# Stimulation of Enzyme Reaction Rates by Crystalline Substrate Irradiation: Dependence on Identity of Irradiated Substance

George E. Bass and James E. Chenevey (deceased)
5567 Ackerman Cove, Bartlett, TN 38134, gbass@uthsc.edu

**Abstract**

The study reported here concerns a phenomenon, discovered and extensively investigated by Sorin Comorosan, wherein enzyme initial reaction rates are enhanced as a consequence of incorporation of solutions derived from previously irradiated crystalline material into the reaction medium. Effective irradiation times conform to a sharply oscillatory pattern. In most reports, the irradiated crystalline material has been the substrate for the enzyme reaction to be studied, but there have been exceptions. The experiments presented here serve to confirm and extend this latter aspect of the phenomenon. It is found that the initial reaction rates for the lactic acid dehydrogenase (LDH) conversion of pyruvate to lactate can be stimulated by irradiation of crystalline deposits of sodium chloride, sodium bromide, potassium chloride and diatomaceous earth. Similarly, stimulation of the LDH conversion of lactate to pyruvate is demonstrated for irradiated sodium chloride. There appears to be no required chemical feature of the irradiated material other than crystalline state.

**Introduction**

Beginning in the late 1960s, S. Comorosan and co-workers[1,2,3,4,5] began reporting an unusual phenomenon wherein an enzyme substrate, in the crystalline state, irradiated for a fixed number of seconds with visible light could lead to a measurable enhancement of the subsequently measured enzyme reaction rate. The irradiation times, $t^*$, that produced this effect were found to be oscillatory and could be represented as $t^* = t_m + n\tau$, where $t_m$ is the shortest time found to produce the effect, n is an integer (0, 1, 2…) and $\tau$ is the constant period separating subsequent enhancements. Over 50 enzyme reactions have been found to demonstrate this effect[6]. All $t^*$ have been found to be a multiple of 5.0 (±<0.5) seconds. Values of $t_m$ range from 5.0 seconds to 45.0 seconds. Values of $\tau$ range from 15.0 seconds to 35.0 seconds. The maximum value of the integer n has varied between laboratories.

In the first report[1] of experiments involving this phenomenon, Comorosan's attention was focused on the enzyme substrate as the entity whose quantized "biological observables" would be perturbed by irradiation and revealed by subsequent enzyme reaction rates. Thus, it was the purified crystalline substrates (urea, sodium isocitrate, potassium malate and sodium glutamate) which were irradiated prior to measurements of their enzyme reaction rates (urease, isocitrate dehydrogenase, malic dehydrogenase and glutamate dehydrogenase, respectively). However, in two following studies[2,4], he found that a non-reactant (cytidine or thymine) could be irradiated, mixed with the crystalline substrate (glucose-6-phosphate or xanthine, respectively) and achieve the same enhanced enzyme reaction rate (glucose-6-phosphate dehydrogenase or xanthine oxidase, respectively). His interpretation at the time was that the irradiated non-reactant crystals transferred the perturbation to the substrate crystals. This facet of the phenomenon was not touched on again until a report[7] in 1980 which detailed experiments wherein irradiated crystalline sodium chloride was effective in stimulating the glutamate-pyruvate transaminase (GPT) conversion of alanine to pyruvate. Comorosan suggested that the irradiated crystals transfer the perturbation to water molecules which then interact with the enzyme such as to produce a higher reaction rate. In 1988, Comorosan[8] suggested an exciton based model wherein a "phonon wind" would modify the energetic topological configuration of the crystal lattice in a manner consistent with any type of crystal.

Uncovering the mechanism whereby this phenomenon is manifest will require a better understanding of the requirements, or limitations, on the identity of the irradiated material. In the series of experiments presented here, the LDH / pyruvate reaction and the LDH / lactate reaction are investigated using irradiated crystalline sodium chloride, potassium chloride, sodium bromide, and diatomaceous earth (insoluble, primarily silica).

**Materials and Methods**

All crystalline samples for irradiation were prepared in Falcon 3001, 35x10 mm cell culture dishes by vacuum (water aspiration) evaporation of 0.250 ml of a double-distilled water solution containing a concentration of the material which provided the desired quantity of deposit, specifically: sodium pyruvate, 3mg; lithium lactate, 15 mg; sodium chloride, 1.6 mg for the pyruvate reaction and 9.13 mg for the lactate reaction; sodium bromide, 2.81 mg; potassium chloride, 2.03 mg; diatomaceous earth, 5.00 mg and 10.0 mg (used for filtered post-irradiation suspension). For enzyme reaction rate measurements, the crystalline deposits were dissolved (or slurried, for diatomaceous earth) in 1.00 ml water and 0.100 ml transferred to a 1 cm spectrophotometer cuvette containing 2.75 ml of NADH (for the pyruvate reaction, 0.20 mg/ml of pH 7.4 phosphate buffer) or NAD (for the lactate reaction, 0.69 mg/ml of pH 7.4 phosphate buffer). The reaction was initiated in the spectrophotometer by addition of 0.050 ml of LDH solution (40 units/ml for pyruvate reaction, 50 units/ml for lactate reaction (rabbit muscle LDH, type II, Sigma Chemical Co.). Change in absorbance for the first 12 seconds of reaction, $\Delta A_{340}$, was determined from the spectrophotometer chart recording.

Crystalline samples were irradiated for specific times with a narrow band selected from the output of a high pressure mercury lamp (General Electric H 100 A4/T, $\lambda_{max}$ = 546.1 nm, bandwidth = 8.8 nm, Detric Optics 2-cavity bandpass filter). The distance of the lamp above the sample (approximately 19 cm) had been adjusted to produce an illuminance of 400 – 500 footcandles (Panlux Electronic Footcandle Meter) at the sample.

To take into consideration any variations in room temperature and to offset possible fluctuations in LDH, NADH and NAD stock solution activities over the course of a day's experiments (and from one enzyme solution batch to another) the crystalline samples were assayed in groups of four, at least one of which was always a non-irradiated control. In the following, each of these groups of four samples is referred to as a "Run." In every Run, assay of the sample(s) irradiated for a t* time was always preceded and followed by assay of samples which were not irradiated or which were irradiated for non-t* times (also considered "control" samples). This allows distinction between irradiation-induced activation and possible activity drift of one sort or another in the course of a Run. In a given Run, all samples were first irradiated for the times indicated, then all were dissolved and loaded into cuvettes as described above. Following addition of NADH or NAD solution to all cuvettes, the four samples were assayed enzymatically within a lapse time of 10 min. The rapid manual addition and mixing of the enzyme was accomplished in 1.5 to 2.0 sec. Lapse time from beginning of the crystalline lactate dissolution step to completion of the enzyme assays was always less than 30 min. Dissolution of the samples was usually initiated within 5 – 10 minutes after the irradiation step, always within thirty minutes. The crystalline lithium lactate samples were always used within 36 hours following their preparation.

**Results**

Irradiation of crystalline sodium chloride was found to yield enhanced rates, relative to non-irradiated controls, for the rabbit muscle LDH / pyruvate reaction for irradiation times of 5, 35, 155 and 995 seconds, corresponding to $t_m$ = 5 seconds, $\tau$ = 30 secconds and n=0, 1, 5 and 33. No enhancement was found for irradiation times of 4, 6, 10, 15, 20, 25, 30, 34, and 36 seconds. These findings are identical to those reported previously[9] for irradiated crystalline sodium pyruvate. Representative data are presented in Table 1. For the rabbit muscle LDH / lactate reaction, irradiated crystalline sodium chloride yielded enhanced reaction rates for irradiation times of 15, 45, 165 and 975 seconds, corresponding to $t_m$ = 15 seconds, $\tau$ = 30 and n=0, 1, 5 and 32. Again, these results are equivalent to those previously reported[9] for irradiation of crystalline lithium lactate. Representative results are presented in Table 2.

Table 1. Sodium Chloride Irradiation Stimulation of the Pyruvate – Lactic Dehydrogenase^ Reaction Rate

|  | $\Delta A_{340}$ / 12-sec enzyme reaction‡ | | | | | $\Delta(\Delta A_{340})$ |
|---|---|---|---|---|---|---|
| Column Label | A | B | C | D | E | F |
|  |  |  |  |  |  |  |
| Run # | t† | 0 sec | t - 1 sec | t sec | t + 1 sec |  |
| 1 | 5 | 0.330 | 0.331 | 0.339 | 0.331 | 0.008 |
| 2 | 5 | 0.321 | 0.322 | 0.333 | 0.321 | 0.012 |
| 3 | 5 | 0.320 | 0.321 | 0.329 | 0.320 | 0.009 |
|  |  |  |  |  |  |  |
| 4 | 35 | 0.328 | 0.329 | 0.338 | 0.329 | 0.009 |
| 5 | 35 | 0.315 | 0.315 | 0.324 | 0.315 | 0.009 |
| 6 | 35 | 0.316 | 0.315 | 0.324 | 0.315 | 0.009 |
|  |  |  |  |  |  |  |
| 7 | 155 | 0.348 | 0.348 | 0.355 | 0.347 | 0.007 |
| 8 | 155 | 0.347 | 0.347 | 0.356 | 0.348 | 0.009 |
| 9 | 155 | 0.348 | 0.348 | 0.356 | 0.347 | 0.008 |
| | | | | | | |
| | | 0 sec | t sec | t sec | 0 sec | |
| 10 | 5 | 0.364 | 0.375 | 0.376 | 0.362 | 0.013 |
| 11 | 10 | 0.368 | 0.367 | 0.368 | 0.368 | -0.001 |
| 12 | 15 | 0.366 | 0.365 | 0.365 | 0.365 | 0.000 |
| 13 | 20 | 0.363 | 0.364 | 0.364 | 0.364 | +0.001 |
| 14 | 25 | 0.366 | 0.367 | 0.366 | 0.367 | 0.000 |
| 15 | 30 | 0.367 | 0.366 | 0.365 | 0.366 | -0.001 |
| 16 | 35 | 0.365 | 0.375 | 0.374 | 0.365 | 0.010 |
| | | | | | | |
| 17 | 995 | 0.351 | 0.359 | 0.360 | 0.352 | 0.008 |
| 18 | 995 | 0.350 | 0360 | 0.359 | 0.350 | 0.009 |
| 19 | 995 | 0.349 | 0.359 | 0.359 | 0.350 | 0.009 |
| | | | | | | |
| Summary: $t_m$ = 5 sec, $\tau$ = 30 sec | | | | | | |

^Rabbit muscle enzyme.
‡Change in absorbance at 340nm in the first 12 seconds of the enzyme reaction.
† Crystalline sodium chloride irradiation time, seconds.
˜ $\Delta(\Delta A_{340})$ = average irradiation induced increase in $\Delta A_{340}$ = **(D-(B+C+D)/3)** for Runs 1-9 and **((C+D)-(B+F))/2** for Runs 10-19, where **B, C, D, F** are the Column Labels.

Table 2. Sodium Chloride Irradiation Stimulation of the Lactate – Lactic Dehydrogenase^ Reaction Rate

| | $\Delta A_{340}$ / 12-sec enzyme reaction‡ | | | | | $\Delta(\Delta A_{340})$˜ |
|---|---|---|---|---|---|---|
| **Column Label** | **A** | **B** | **C** | **D** | **E** | **F** |
| | | | | | | |
| **Run #** | $t^†$ | 0 sec | t - 1 sec | t sec | t + 1 sec | |
| 1 | 15 | 0.0720 | 0.0721 | 0.0737 | 0.0722 | 0.0016 |
| 2 | 15 | 0.0725 | 0.0724 | 0.0740 | 0.0724 | 0.0016 |
| 3 | 15 | 0.0726 | 0.0727 | 0.0741 | 0.0726 | 0.0015 |
| | | | | | | |
| 4 | 45 | 0.0702 | 0.0701 | 0.0718 | 0.0702 | 0.0016 |
| 5 | 45 | 0.0700 | 0.0702 | 0.0721 | 0.0703 | 0.0019 |
| 6 | 45 | 0.0702 | 0.0703 | 0.0719 | 0.0701 | 0.0017 |
| | | | | | | |
| 7 | 165 | 0.0732 | 0.0733 | 0.0749 | 0.0732 | 0.0017 |
| 8 | 165 | 0.0731 | 0.0730 | 0.0746 | 0.0730 | 0.0016 |
| 9 | 165 | 0.0733 | 0.0733 | 0.0748 | 0.0732 | 0.0015 |
| | | | | | | |
| 10 | 975 | 0.0724 | 0.0723 | 0.0740 | 0.0725 | 0.0016 |
| 11 | 975 | 0.0722 | 0.0723 | 0.0738 | 0.0722 | 0.0017 |
| 12 | 975 | 0.0721 | 0.0720 | 0.0737 | 0.0722 | 0.0016 |
| | | | | | | |
| | | 0 sec | t sec | t sec | 0 sec | |
| 13 | 5 | 0.0740 | 0.0739 | 0.0740 | 0.0741 | -0.0001 |
| 14 | 10 | 0.0738 | 0.0740 | 0.0739 | 0.0740 | +0.0001 |
| 15 | 20 | 0.0739 | 0.0739 | 0.0739 | 0.0740 | -0.0001 |
| 16 | 25 | 0.0742 | 0.0741 | 0.0743 | 0.0743 | -0.0001 |
| 17 | 30 | 0/0735 | 0.0737 | 0.0736 | 0.0736 | +0.0001 |
| 18 | 35 | 0.0734 | 0.0734 | 0.0736 | 0.0735 | +0.0001 |
| 19 | 40 | 0.0733 | 0/0732 | 0/0733 | 0.0734 | -0.0001 |
| | | | | | | |
| | Summary: $t_m$ = 15 sec, $\tau$ = 30 sec | | | | | |

^Rabbit muscle enzyme.
‡Change in absorbance at 340nm in the first 12 seconds of the enzyme reaction.
† Crystalline sodium chloride irradiation time, seconds.
~ $\Delta(\Delta A_{340})$ = average irradiation induced increase in $\Delta A_{340}$ = **(D-(B+C+D)/3)** for Runs 1-12 and **((C+D)-(B+F))/2** for Runs 13-19, where **B, C, D, F** are the Column Labels.

Similarly, irradiations of crystalline potassium chloride, sodium bromide and diatomaceous earth were found to yield enhanced reaction rates for the pyruvate / LDH reaction for irradiation times of 5 and 35 seconds. These correspond to : $t_m$ = 5 seconds, $\tau$ = 30 seconds and n=0 and 1. These are equivalent to the results presented above for sodium chloride and reported previously for irradiation of crystalline sodium pyruvate. Representative results are presented in Tables 3 – 7.

Table 3. Potassium Chloride Irradiation Stimulation of the Pyruvate – Lactic Dehydrogenase^ Reaction Rate

| | $\Delta A_{340}$ / 12-sec enzyme reaction‡ | | | | | $\Delta(\Delta A_{340})$ ~ |
|---|---|---|---|---|---|---|
| **Column Label** | **A** | **B** | **C** | **D** | **E** | **F** |
| | | | | | | |
| **Run #** | t† | 0 sec | t sec | t sec | 0 sec | |
| 1 | 5 | 0.355 | 0.364 | 0.365 | 0.356 | 0.009 |
| 2 | 5 | 0.354 | 0.364 | 0.363 | 0.354 | 0.009 |
| 3 | 5 | 0.355 | 0.363 | 0.363 | 0.355 | 0.008 |
| | | | | | | |
| 4 | 35 | 0.351 | 0.360 | 0.361 | 0.352 | 0.009 |
| 5 | 35 | 0.350 | 0.358 | 0.359 | 0.350 | 0.009 |
| 6 | 35 | 0.351 | 0.359 | 0.358 | 0.350 | 0.007 |
| | | | | | | |
| Summary: $t_m$ = 5 sec, $\tau$ = 30 sec | | | | | | |

^Rabbit muscle enzyme.

‡Change in absorbance at 340nm in the first 12 seconds of the enzyme reaction.
† Crystalline potassium chloride irradiation time, seconds.
~ $\Delta(\Delta A_{340})$ = average irradiation induced increase in $\Delta A_{340} = ((C+D)-(B+F))/2$, where B, C, D, F are the Column Labels.

Table 4.  Sodium Bromide Irradiation Stimulation of the Pyruvate – Lactic Dehydrogenase^ Reaction Rate

|  | $\Delta A_{340}$ / 12-sec enzyme reaction‡ | | | | | $\Delta(\Delta A_{340})$~ |
|  | A | B | C | D | E | F |
| Run # | t† | 0 sec | t sec | t sec | 0 sec |  |
| 1 | 5 | 0.344 | 0.352 | 0.353 | 0.345 | 0.008 |
| 2 | 5 | 0.343 | 0.353 | 0.352 | 0.344 | 0.009 |
| 3 | 5 | 0.342 | 0.351 | 0.350 | 0.343 | 0.008 |
|  |  |  |  |  |  |  |
| 4 | 35 | 0.341 | 0.349 | 0.349 | 0.340 | 0.009 |
| 5 | 35 | 0.342 | 0.350 | 0.350 | 0.342 | 0.008 |
| 6 | 35 | 0.340 | 0.348 | 0.349 | 0.340 | 0.008 |
|  | Summary:  $t_m$ = 5 sec,  $\tau$ = 30 sec | | | | | |

^Rabbit muscle enzyme.
‡Change in absorbance at 340nm in the first 12 seconds of the enzyme reaction.
† Crystalline sodium bromide irradiation time, secseconds.
~ $\Delta(\Delta A_{340})$ = average irradiation induced increase in $\Delta A_{340} = ((C+D)-(B+F))/2$, where B, C, D, F are the Column Labels.

Table 5.  Diatomaceous Earth Irradiation Stimulation of the Pyruvate – Lactic Dehydrogenase^ Reaction Rate:  Unfiltered Suspension Added to Reaction Mixture.

|  | $\Delta A_{340}$ / 12-sec enzyme reaction‡ | | | | | $\Delta(\Delta A_{340})$~ |
| Column Label | A | B | C | D | E | F |
| Run # | t† | 0 sec | t sec | t sec | 0 sec |  |
| 1 | 5 | 0.291 | 0.300 | 0.299 | 0.292 | 0.008 |
| 2 | 5 | 0.292 | 0.301 | 0.301 | 0.292 | 0.009 |
| 3 | 5 | 0.291 | 0.299 | 0.299 | 0.290 | 0.008 |
|  |  |  |  |  |  |  |
| 4 | 35 | 0.290 | 0.299 | 0.299 | 0.291 | 0.009 |
| 5 | 35 | 0.291 | 0.300 | 0.298 | 0.290 | 0.008 |
| 6 | 35 | 0.289 | 0.298 | 0.299 | 0.290 | 0.009 |

| | | Summary: $t_m$ = 5 sec, $\tau$ = 30 sec | | | |

^Rabbit muscle enzyme.
‡Change in absorbance at 340nm in the first 12 seconds of the enzyme reaction.
† Crystalline diatomaceous earth irradiation time, seconds.
~ $\Delta(\Delta A_{340})$ = average irradiation induced increase in $\Delta A_{340}$ = ((C+D)-(B+F))/2, where B, C, D, F are the Column Labels.

Table 6. Diatomaceous Earth Irradiation Stimulation of the Pyruvate – Lactic Dehydrogenase^ Reaction Rate: Filtered# Solution Added to Reaction Mixture.

| | | $\Delta A_{340}$ / 12-sec enzyme reaction‡ | | | | $\Delta(\Delta A_{340})$ ~ |
|---|---|---|---|---|---|---|
| **Column Label** | A | B | C | D | E | F |
| **Run #** | t† | 0 sec | t sec | t sec | 0 sec | |
| 1 | 5 | 0.287 | 0.295 | 0.296 | 0.288 | 0.008 |
| 2 | 5 | 0.289 | 0.296 | 0.296 | 0.289 | 0.007 |
| 3 | 5 | 0.288 | 0.296 | 0.295 | 0.288 | 0.008 |
| | | | | | | |
| 4 | 35 | 0.287 | 0.295 | 0.294 | 0.287 | 0.007 |
| 5 | 35 | 0.287 | 0.294 | 0.295 | 0.287 | 0.008 |
| 6 | 35 | 0.288 | 0.296 | 0.295 | 0.288 | 0.007 |
| | | | | | | |
| Summary: $t_m$ = 5 sec, $\tau$ = 30 sec | | | | | | |

^Rabbit muscle enzyme.
# Post-irradiation filtration, 0.22 micrometer Millipore filter.
‡Change in absorbance at 340nm in the first 12 seconds of the enzyme reaction for crystalline samples irradiated for **0** or **t** seconds..
† Crystalline diatomaceous earth irradiation time, seconds.
~ $\Delta(\Delta A_{340})$ = average irradiation induced increase in $\Delta A_{340}$ = ((C+D)-(B+F))/2, where B, C, D, F are the Column Labels.

Several experiments were conducted in which empty cell culture dishes (in which the crystalline samples were prepared and irradiated) were irradiated for t*-times and then loaded with the employed volume and concentrations of salt solutions. In all such tests, no effect on enzyme reaction rate was found.

**Discussion**

The results presented here serve to confirm and extend Comorosan's finding that enzyme reaction rate enhancements can be achieved by irradiation of crystalline materials other than the enzyme substrate. However, contrary to Comorosan's early views, there is no unique photochemical role played by the crystalline enzyme substrate, *per se*, in achieving the effect. Further, there appears to be no particular chemical compositional feature, beyond provision of a crystalline state, that is required. The smooth plastic surface provided by the bottom of the cell culture dish was found to be inadequate for this purpose.  Beyond the irradiation apparatus, the only apparent features common to the entire body of work that currently defines this phenomenon are that (1) the irradiations are performed in the open atmosphere of the laboratory and (2) a crystalline material is present.  Thus, we are led to propose that the occurrence of this phenomenon derives from photo-driven oscillatory reactions among molecules of the atmosphere, at or near the crystalline surface, that lead to deposition of chemical species on the crystalline material, which when dissolved in water, are capable of enhancing the initial reaction rates of a very wide variety of enzymes and influencing growth rates of living microorganisms.